\documentclass{iau}

\usepackage{amsmath}
\usepackage{graphicx}
\usepackage{multirow}
\usepackage{orcidlink}

\begin{document}

\lefttitle{D. Calderón}
\righttitle{Stellar winds feeding Sgr~A*}

\jnlPage{1}{7}
\jnlDoiYr{2026}
\doival{10.1017/xxxxx}

\aopheadtitle{Proceedings IAU Symposium}
\editors{M. Zaja\v{c}ek,  T. Je\v{r}\'{a}bkov\'{a}, V. Karas, R. Schödel \&  P. Sukov\'{a}, eds.}

\title{Interacting stellar winds feeding Sgr A*: from the system of mass-losing stars to the binary IRS 16SW}

\author{Diego Calderón$^{1}$\footnote{Alexander von Humboldt Fellow}\orcidlink{0000-0002-9019-9951}, Jorge Cuadra$^{2,3}$\orcidlink{0000-0003-1965-3346}, Christopher M. P. Russell$^4$\orcidlink{0000-0002-9213-0763}, Andreas Burkert$^{5,6}$\orcidlink{0000-0001-6879-9822}, Stephan Rosswog$^{7,8}$\orcidlink{0000-0002-3833-8520}, Mayura Balakrishnan$^9$\orcidlink{0000-0001-9641-6550}}
\affiliation{$^1$Max-Planck-Institut für Astrophysik, Karl-Schwarzschild-Straße 1, 85748 Garching, Germany
\email{calderon@mpa-garching.mpg.de}}
\affiliation{$^2$Universidad Adolfo Ib\'a\~nez, Av. Padre Hurtado 750, Vi\~na del Mar, Chile}
\affiliation{$^3$Millennium Nucleus on Transversal Research and Technology to Explore Supermassive Black Holes (TITANS), Chile}
\affiliation{$^4$Department of Physics and Astronomy, Bartol Research Institute, University of Delaware, Newark, DE 19716, USA.}
\affiliation{$^5$Universitäts-Sternwarte, Ludwig-Maximilians-Universität München, Scheinerstr. 1, 81679 Munich, Germany}
\affiliation{$^6$Max Planck Institute for Extraterrestrial Physics, Giessenbacherstr. 1, 85748 Garching, Germany}
\affiliation{$^7$Hamburger Sternwarte, Universität Hamburg, Gojenbergsweg 112, 21029 Hamburg, Germany}
\affiliation{$^8$The Oskar Klein Centre, Department of Astronomy, AlbaNova, Stockholm University, 106 91 Stockholm, Sweden}
\affiliation{$^9$Trottier Space Institute at McGill University 3550 Rue University, H3A 0C6, Montreal, QC, Canada}

\begin{abstract}
    The discovery of cold structures around Sgr A* has challenged our understanding of the gas dynamics and thermodynamic state of the plasma in its vicinity. This work aims to constrain the conditions for the formation of such structures, namely the cold disc-like structure and the recently discovered G-1-2-3 complex. We conduct hydrodynamic simulations of the observed Wolf-Rayet stars feeding Sgr A*. Our simulations show that the plasma chemical composition is crucial for determining the medium properties. We demonstrate that the formation of a cold disc is possible for chemical compositions that are consistent with observational constraints. However, it is not possible to reproduce all the properties of the observed disc which might suggest the interaction with another structure. Additionally, we present our first results on the hydrodynamic modelling of IRS 16SW as a colliding-wind binary. This is the first step to develop a realistic model on the formation of the G-1-2-3 complex.
\end{abstract}

\begin{keywords}
    Galaxy: centre – ISM: clouds – Stars: winds, outflows – Stars: binaries – Stars: Wolf-Rayet
\end{keywords}

\maketitle

\section{Introduction}

    The central parsec of the Milky Way hosts the super-massive black hole Sagittarius~A* (Sgr~A*). 
    Its vicinity is inhabited by millions of stars which correspond to the central nuclear star cluster. 
    Among them, there is an unusual high number of massive and young stars in the region \citep[e.g.][]{paumard2006}. 
    Some of them are in the Wolf-Rayet phase which is characterised for having strong outflows in form of stellar winds accelerated at high speeds \citep[e.g.][]{martins2007}. 
    It is thought that these mass-losing stars are the main responsible for feeding Sgr~A* as numerical modelling has shown that the inflow mass rate is consistent with observational constraints \citep[e.g.][]{cuadra2008,ressler2018,calderon2020b}. 
    Yet the state of the medium is complex and resembles multi-phase components. 
    Cold and dusty structures inhabit the central parsec coexisting with the hot, diffuse plasma of the shocked stellar winds. 
    For instance, observations of the H30$\alpha$ emission line at the position of Sgr~A* indicate the presence of a disc-like structure around it \citep{murchikova2019}. 
    In addition, the dusty, gaseous object G2, whose monitoring extends for over a decade \citep[e.g.][]{gillessen2012,peissker2025}, has been shown to be part of a larger streamer of gas containing similar sources such as G1 and G3 \citep{gillessen2026}. 
    The dynamics of this so-called G-1-2-3 streamer points to an origin related to the massive binary IRS 16SW.

\section{Numerical Simulations}

    We performed numerical simulations using the adaptive-mesh refinement hydrodynamic code Ramses \citep{teyssier2002}. 
    The code solves the hydrodynamic equations in their Eulerian form and has a module to enhance the spatial resolution based on physical criteria.    
    In this work, we studied two different setups: $i)$ a hydrodynamic modelling of 30 Wolf-Rayet stars feeding Sgr~A* \citep{calderon2025}, and $ii)$ a hydrodynamic modelling for IRS 16SW interpreted as a colliding-wind binary (Calderon et al., in prep.). 

    \noindent
    $i)$ 
    This model considered a cubic domain of side 1.6~pc where the stars move under the gravitational potential of Sgr~A* considered as a point mass of $4.3\times10^6$~M$_{\odot}$ placed at the centre of the domain. 
    The stars are initialised so that they move on their constrained observed orbits \citep[e.g.][]{paumard2006}. 
    In addition, the stars act as sources of mass, momentum, and energy in the form of stellar winds whose properties are known from spectroscopic studies \citep{martins2007}. 
    The optically-thin radiative cooling was determined using pre-computed tabulated emissivities \citep{stofanova2021} which were adapted for the sub-type of Wolf-Rayet stars. 
    The simulation starting point was computed extrapolating the current star positions into the past, so that the simulation covered 3,500~yr reaching the present time. 
    For further details, we defer the reader to \cite{calderon2025}.

    \noindent 
    $ii)$
    This simulation considered a cubic domain of side 2~au where two identical stars were placed in a binary system. 
    The orbital properties of the binary system were taken from \cite{martins2006}: a period of $\sim$19.5~d, stellar separation of $\sim$0.7~au, and stellar masses of 50~M$_{\odot}$ each. 
    The stellar radii were chosen to be 40~R$_{\odot}$, so that no component filled its Roche lobe. 
    Each star blows a steady spherical wind with mass-loss rates and terminal velocities of $1.12\times10^{-5}$~M$_{\odot}$~yr$^{-1}$ and 600~km~s$^{-1}$, respectively \citep{martins2007,cuadra2008}.
    Given the short stellar binary separation, it was necessary to take into account the stellar wind acceleration as the winds might not reach terminal speed before they collide. 
    To do so, we used the ``anti-gravity" approach so that the stellar winds follow the empirical $\beta$ velocity laws using $\beta=1$. 
    The simulation was run for a total of four orbital periods (i.e. $\sim$80~d).

\section{Results}
    \subsection{System of mass-losing stars}
\begin{figure}
            \centering
            \includegraphics[width=0.5\linewidth]{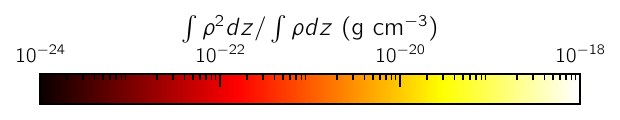}
            \\
            \includegraphics[width=0.45\linewidth]{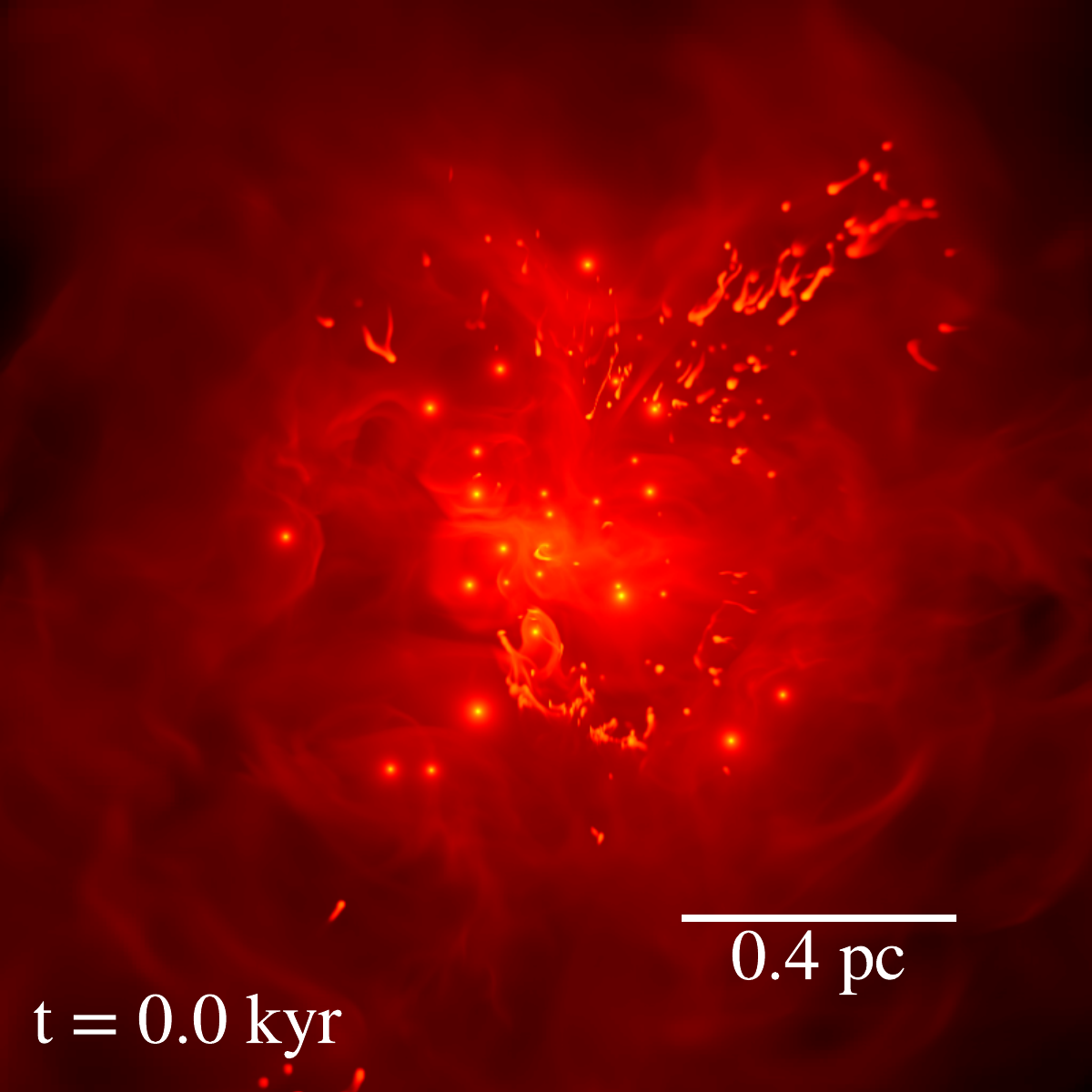}
            \includegraphics[width=0.45\linewidth]{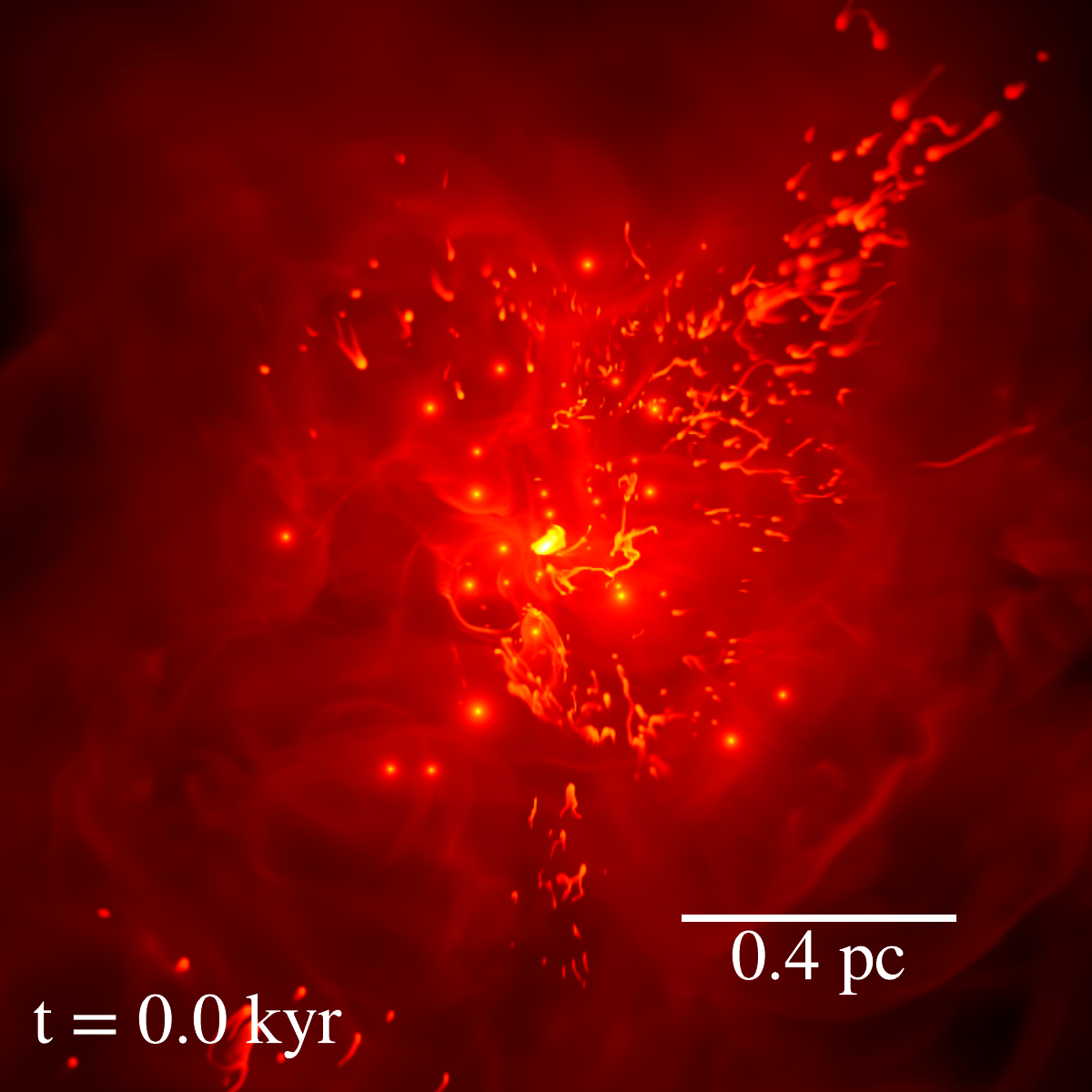}
            \caption{
            Density maps of the simulations of the system of mass-losing stars feeding Sgr~A*.
            The panels show projected density maps weighed by density along the $z$ axis, i.e. $\int \rho^2dz/\int\rho dz$, which is parallel to the line of sight. 
            The left- and right-hand side panels show the fiducial and enhanced-cooling setups, respectively. 
            Both panels show the whole domain at the final state that corresponds to the present time.
            }
            \label{fig:gc}
        \end{figure}

        The final state of the simulation of the system of Wolf-Rayet stars feeding Sgr~A* via stellar winds is shown in Figure~\ref{fig:gc}. 
        Notice that this simulation time corresponds to the system in the present time. 
        The left- and right-hand side panels show the outcome of two different models: the fiducial and enhanced radiative cooling runs, respectively. 
        The fiducial case used the chemical abundances reported by \cite{russell2017} to compute the optically-thin radiative cooling emissivities. 
        The enhanced radiative cooling run used different chemical abundances that maximised the emissivities but are still consistent with the observational constraints \cite[see][for a discussion]{calderon2025}.
        As a result, the line-of-sight integrated density (weighed by density) maps show analogous configurations globally but differ significantly on the amount of clumpy structures and accumulation at the location of Sgr~A* at the centre of the domain (see Figure~\ref{fig:gc}).

        The fiducial case shows diffuse gas filling the whole domain with higher densities toward the centre, as expected. 
        In this case, a moderate amount of denser (and cooler) structures can be observed especially in the upper right region of the domain. 
        These clumps are the result of radiative cooling together with hydrodynamic instabilities that develop in the bow shock of the stellar winds, and are pushed away from the centre due to the net effect of the stellar winds that overall outflow.
        In the case of the enhanced radiative cooling run, the clumpy structures are more common also on the upper right region of the domain. 
        However, more importantly there is a clear accumulation of material around Sgr~A*. 
        This structure corresponds to a cold ($\sim$10$^4$~K), dense disc whose total mass is about 5$\times$10$^{-3}$~M$_{\odot}$ and a diameter of $\sim$0.04~pc (or $\sim$1~arcsec). 
        The presence of this disc is a direct consequence of the more efficient radiative cooling that results into more cold, dense clumps that manage to infall onto the central region. 
        Although not shown here, this fact can also be observed on the mass inflow rate across the central boundary whose amplitude is two to four times higher than in the fiducial case. 

        The formation of a cold disc observed in our hydrodynamic simulation could be a natural explanation for the reported disc-like structure around Sgr~A* \citep{murchikova2019}. 
        To make a quantitative comparison, we performed further analysis on the properties of the simulated disc. 
        Specifically, we analysed the line-of-sight velocity structure of the disc and synthesised fluxes of the Br$\gamma$ and H30$\alpha$ recombination lines. 
        The simulated disc shows a red- and blueshifted structure spanning from its left- to right-hand sides, respectively. 
        The velocity is within the range [-2000:2000]~km~s$^{-1}$, in agreement with the observations. 
        However, the inclination of the simulated disc is not exactly the same as the observed one. 
        There is a noticeable difference of about $90^{\circ}$. 
        Regarding the recombination line fluxes, the simulated fluxes are not simultaneously consistent with the observational constraints. 
        Unfortunately, the fiducial model is consistent with the Br$\gamma$ upper limit \citep{ciurlo2021} but not with the observed H30$\alpha$ flux \citep{murchikova2019}. 
        Vice-versa, the enhanced cooling model is not consistent with the Br$\gamma$ upper limit but in agreement with the reported H30$\alpha$ flux. 
        We speculate that such inconsistencies could be a reflection of the lack of certain known structures that our numerical models do not consider such as the circumnuclear disc \citep[e.g.][]{solanki2023}.

    \subsection{IRS 16SW}

\begin{figure}
            \includegraphics[width=0.425\linewidth]{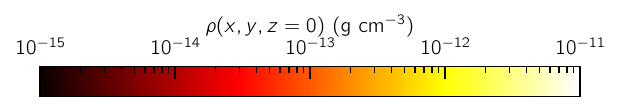}
            \hfill
            \includegraphics[width=0.425\linewidth]{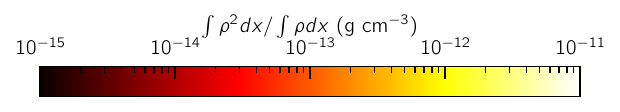}
            \includegraphics[width=0.425\linewidth]{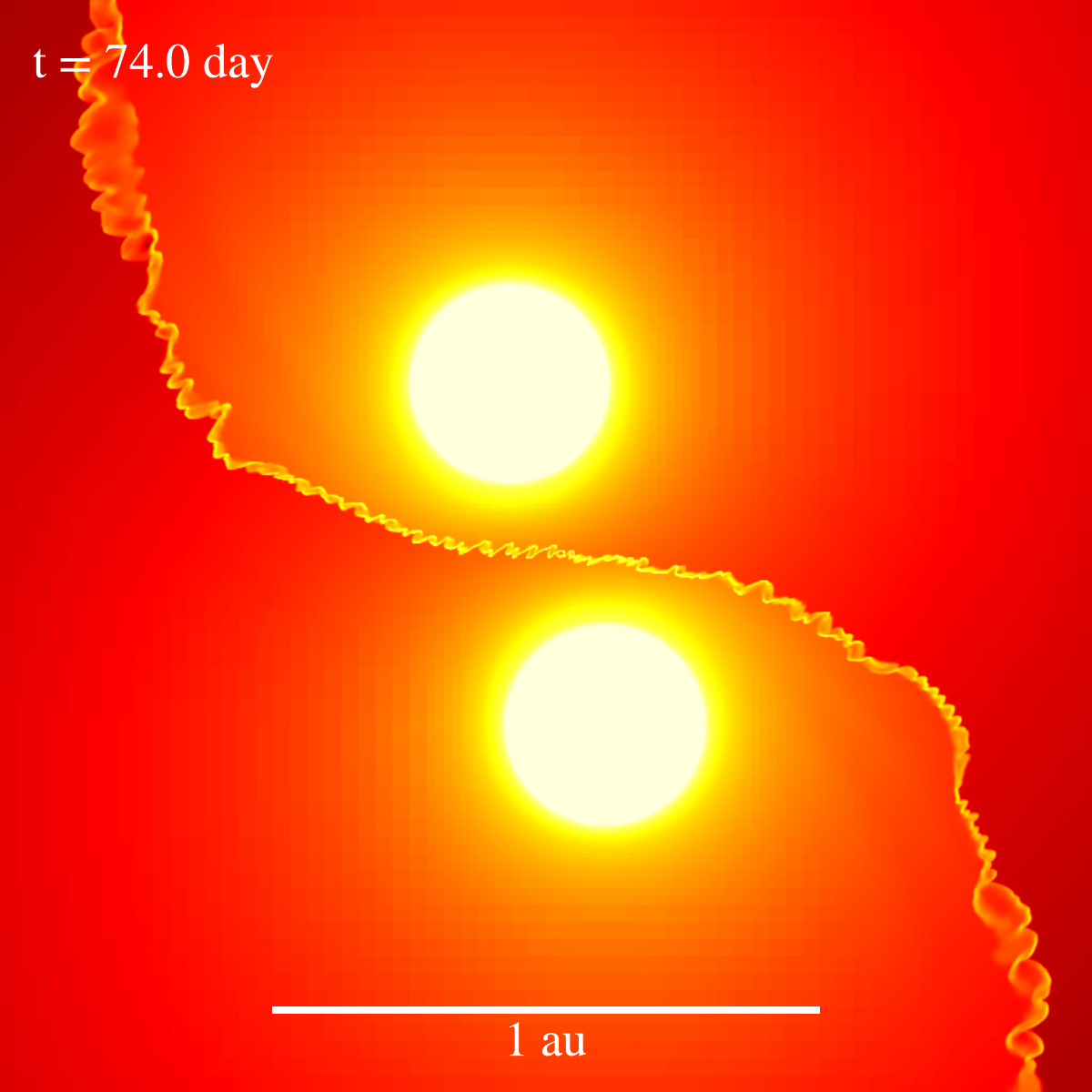}
            \hfill
            \includegraphics[width=0.425\linewidth]{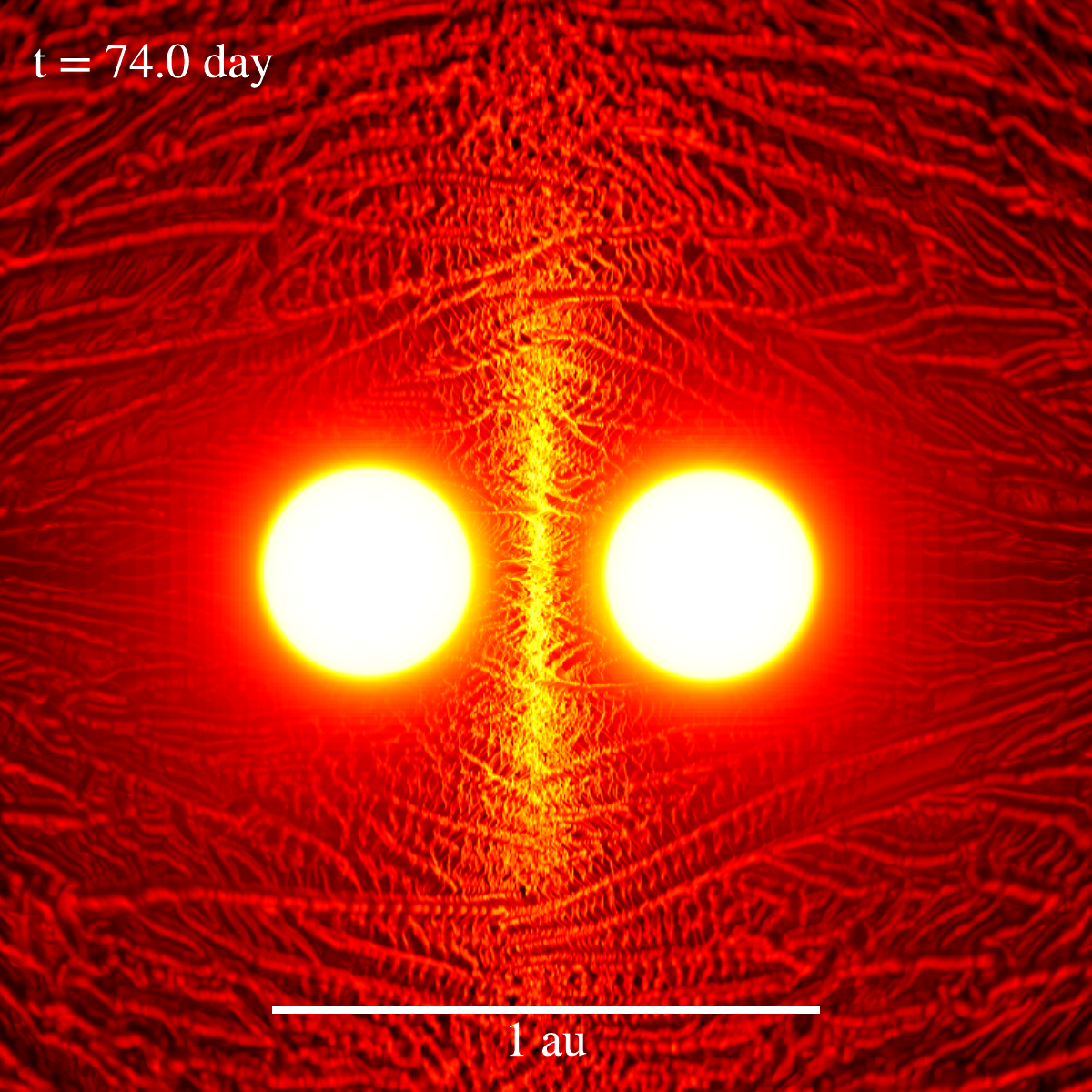}
            \caption{
            Density maps of the IRS~16SW as a colliding-wind binary.  
            The left- and right-hand side panels show a slice on the orbital plane $z=0$ and a density-weighed integration along the $x$ axis, i.e. $\int \rho^2dx/\int\rho dx$, which is parallel to the line of sight, respectively. 
            Both maps show whole domain at the same simulation time $t=74$~d.
            The binary orbital motion is counter clockwise, i.e. +z direction.
            }
            \label{fig:binary}
        \end{figure}

        The colliding-wind binary simulation of IRS 16SW reaches a quasi-stationary state relatively fast. 
        In less than an orbital period, the winds fill the whole domain and create a thin shell at the wind-interaction region between the stars. 
        The slab is prone to thin shells instabilities \citep[e.g.][]{vishniac1994}.
        as shown in the density map on the left-hand side of Figure~\ref{fig:binary}. 
        However, the shell has a much more complex structure when analysed in three dimensions. 
        The right-hand side of Figure~\ref{fig:binary} shows a line-of-sight integrated density map (weighed by density) at the time of maximum projected stellar separation. 

        We analysed the slab to assess quantitatively if it can reach a stationary state. 
        To do so, we assigned a scalar tracer field to each stellar wind, so that we can select the cells where both tracers are larger than zero. 
        Once we have selected the slab, we estimated the total mass within the slab as well as its Br$\gamma$ recombination line flux across every single output of the simulation. 
        The results show that the slab contains a total mass of 0.01~M$_{\oplus}$ with a Br$\gamma$ luminosity of 0.5~L$_{\odot}$. 
        These values remain constant throughout the simulation which confirms that the system is in a quasi-stationary state. 
        Although the mass and recombination line flux values are small it seems that the unstable slab could contribute to the observations of IRS~16SW. 

\section{Conclusions}

        We have presented a hydrodynamic study of stellar winds feeding Sgr~A* and of the binary IRS~16SW. 
        The models show that the interaction of the stellar winds fill the central parsec with diffuse material, bow shocks and clumpy structures, depending on the chemical composition of the stellar atmospheres. 
        The current observational constraints are consistent with abundances that allow the formation of a disc around Sgr~A*. 
        In principle, this disc could correspond to the structure reported by \cite{murchikova2019} but the inclination and line fluxes are not fully consistent with the observations. 
        We have also simulated IRS~16SW as a colliding-wind binary. 
        The results show that the system creates a quasi-stationary unstable wind-confined slab whose enclosed mass and Br$\gamma$ luminosity are  0.01~M$_{\oplus}$ and 0.5~L$_{\odot}$, respectively. 
        These properties must be taken into account for interpreting future near-infrared observations. 
        Further modelling of this system in a medium will allow us to assess its role creating the G-1-2-3 complex.

\section*{Acknowledgments}

    DC and JC acknowledge the financial support from ANID-FONDECYT Regular 1251444. 
    The research of DC has been funded by the Alexander von Humboldt Foundation. 
    The numerical simulations of this work were run on the high-performance computing system Raven of the Max Planck Computing and Data Facility. 
    The analysis of the numerical simulation outputs was carried out using the python package \textsc{yt} \citep{turk2011}.

\bibliographystyle{iaulike} 
\bibliography{iaubib}

\end{document}